\documentclass[conference]{IEEEtran}
\IEEEoverridecommandlockouts
\usepackage[numbers]{natbib}
\usepackage{amsmath,amssymb,amsthm,amsfonts}
\usepackage{graphicx}
\usepackage{textcomp}
\usepackage[dvipsnames]{xcolor}
\usepackage{algpseudocode}
\usepackage[linesnumbered,ruled,vlined]{algorithm2e}
\usepackage{tabularx}
\usepackage{todonotes}
\usepackage[colorlinks=true, allcolors=blue]{hyperref}

\newtheorem{defn}{Definition}
\def\BibTeX{{\rm B\kern-.05em{\sc i\kern-.025em b}\kern-.08em
    T\kern-.1667em\lower.7ex\hbox{E}\kern-.125emX}}

\newcommand{\powerps}{p^{(S)}}
\begin{document}

\title{Network Sparsification via Degree- and Subgraph-based Edge Sampling}

\author{
\IEEEauthorblockN{
Zhen Su\IEEEauthorrefmark{1}\IEEEauthorrefmark{2},
J\"{u}rgen Kurths\IEEEauthorrefmark{1}\IEEEauthorrefmark{3}
and
Henning Meyerhenke\IEEEauthorrefmark{2}
}
\IEEEauthorblockA
{
\IEEEauthorrefmark{1}Potsdam Institute for Climate Impact Research, Potsdam, Germany\\
\IEEEauthorrefmark{2}Department of Computer Science, Humboldt-Universität zu Berlin, Berlin, Germany\\
\IEEEauthorrefmark{3}Department of Physics, Humboldt-Universität zu Berlin, Berlin, Germany\\
Email: 
zhen.su@pik-potsdam.de,
kurths@pik-potsdam.de,
meyerhenke@hu-berlin.de}}

\maketitle

\begin{abstract}
Network (or graph) sparsification compresses a graph by removing inessential edges.
By reducing the data volume, it accelerates or even facilitates many downstream analyses.
Still, the accuracy of many sparsification methods, with filtering-based edge sampling being the most typical one, heavily relies on an appropriate definition of edge importance.
Instead, we propose a different perspective with a generalized local-property-based
sampling method, which preserves (scaled) local \emph{node} characteristics.
Apart from degrees, these local node characteristics we use are the expected (scaled) number of wedges and triangles a node belongs to.
Through such a preservation, main complex structural properties are preserved implicitly.
We adapt a game-theoretic framework from uncertain graph sampling by including
a threshold for faster convergence (at least $4$ times faster empirically) to approximate solutions.
Extensive experimental studies on functional climate networks show the effectiveness of this method in preserving macroscopic to mesoscopic and microscopic network structural properties.
\end{abstract}


\section{Introduction}\label{Sec:Introduction}
Network science facilitates the study of various complex systems.
Apart from physically (e.g., technological networks) or conceptually (e.g., social networks) connected entities, time series data from different contexts can be analyzed by relating
nodes to each other that are correlated in some way.
For example, in climate science, by treating locations on earth as nodes and establishing edges between nodes according to the corresponding time series, climate data are represented as networks~\cite{su2022climatic} (= graphs, we use both terms interchangeably).
The resulting objects are often referred to as \emph{functional networks}.

Due to the large size of many real-world networks, downstream analyses, such as visualization and structural queries, can be time-consuming or even prohibitive.
A natural solution often seen in the literature is to discard a large proportion of possibly redundant edges by sparsification (without the aggregation of nodes).
Under the basic premise of preserving essential network properties, it allows 
a faster and sometimes even more accurate analysis of the
available network data~\cite{hamann2016structurepreserving}.

Which properties to preserve with the subgraph resulting from sparsification, depends on the application context.
Theoretical work considered, among others, spectral properties such as eigenvalues~\cite{batson2013spectral}, requiring the solution of many Laplacian linear systems.
For practical applications, alternative objectives that can be computed faster are often preferred.
Typically, this happens by sampling the edges to be preserved in the sparser subgraph $G^*$
according to some probability distribution.
The simplest one is uniform sampling, which preserves a type of restricted spectral property with high probability~\cite{sadhanala2016graph}.
Other sampling methods aiming at preserving structural properties were systematically compared in~\cite{hamann2016structurepreserving}.
The general sampling process used there contains two primary steps: edge scoring and filtering.
Edge scoring assigns each edge a value that describes how `essential' it is; filtering then removes
all edges with scores below a certain threshold such that the network is compressed to a desired
ratio.

By preserving degree- and subgraph-based local properties, one can reconstruct complex 
properties of a given network~\cite{mahadevan2006systematica,orsini2015quantifying}.
This is also true for functional networks. In particular, it has been shown for general networks
that a node's importance correlates with its degree as well as with the number of triangles and wedges
it belongs to~\cite{benzi2015limiting}.
Motivated by this, we want to sparsify the input graph $\mathcal{G}$ such that these three measures above
are preserved in expectation -- just scaled appropriately with the sparsification ratio.

To the best of our knowledge, there is limited work closely related to this objective.
\citet{zeng2021selective,zeng2022reductiona} formulate a similar approach as an optimization
problem, but they preserve only the expected degrees -- which seems overly myopic.
Since triads (connected $3$-node-subgraphs) play an important role in (functional) networks, we thus
transfer results from uncertain graph sampling~\cite{parchas2014pursuit,parchas2015uncertain,song2016trianglebased} to network sparsification.
In uncertain graphs, the objective is to sample a representative instance from the set of
all possible instances. We adapt this idea for sampling edges such that the three desired node properties
above are retained in expectation.
Although the preservation of subgraphs could be extended to larger sizes, the computational cost can be prohibitive above three~\cite{parchas2015uncertain}.

\setlength\extrarowheight{3pt}
\begin{table*}[!t]
  \footnotesize
  \caption{List of symbols.}\label{Tab:Symbols}
  \begin{@twocolumnfalse}
    \setlength\arrayrulewidth{1pt}
    \setlength{\tabcolsep}{3pt}
    \begin{tabularx}{\textwidth}{cX}
      \hline
      Symbol                     & Definition                                                                                                                               \\
      \hline
      $\mathcal{G}=(V, E, p)$    & An undirected network with $|V|$ nodes, $|E|$ edges, and confidence values $p:E\rightarrow (0.95, 1]$ associated with the edges \\
      $\powerps$                      & The generic contribution of each edge to form the final sparse structure, by multiplying a scaling factor $S \in [0, 1] $                 \\
      $G^{*}=(V, E^{*})$         & The final sparse subgraph after edge sampling with $|V|$ nodes, $|E^*|$ edges                                                     \\
      $G'=(V, E')$               & The current subgraph during edge sampling with $|V|$ nodes, $|E'|$ edges                                                               \\
      $l=2,3,w$                  & The basic local properties associated with each node to be preserved, i.e., $l=2$ for degree, $l=3$ for triangles, and/or $l=w$ for wedges  \\
      $m_{l}(u, \mathcal{G})$    & The possible degree of node $u$ and possible number of triangles and wedges $u$ belongs to based on $\mathcal{G}$                                     \\
      $|L_{l}(u, \mathcal{G})|$  & The maximum possible degree of node $u$ and maximum possible number of triangles and wedges $u$ belongs to based on $\mathcal{G}$                                        \\
      $E[m_{l}(u, \mathcal{G})]$ & The expected degree of node $u$ and expected number of triangles and wedges $u$ belongs to based on $\mathcal{G}$                                                \\
      $m_{l}(u, G')$             & The current degree of node $u$ and current number of triangles and wedges $u$ belongs to based on $G'$                                                          \\
      $\Delta m_{l}(u, G')$      & The distance, for node $u$, between the current local properties $m_{l}(u, G')$ based on $G'$ and the corresponding expectations $E[m_{l}(u, \mathcal{G})]$                                 \\
      $\Delta m_{l=2,3,w}(G')$         & The total distance of the current $G'$ to the expectation, summarized over $l=2,3,w$ and over all nodes $V$                              \\
      \hline
    \end{tabularx}
  \end{@twocolumnfalse}
\end{table*}

The contributions of this paper are as follows:
We propose a scaled local-property-based (degrees and 3-node subgraphs) edge sampling, adapted from uncertain graph sampling, for network sparsification.
This new perspective of sparsification relaxes the dependency on a specific edge-scoring method.
{To this end, we adapt a game-theoretic framework~\cite{parchas2015uncertain} and experiments demonstrate that our focus on scaled local properties usually leads to a better preservation of more complex properties than other
state-of-the-art sparsification methods.}

\section{Problem Definition}\label{Sec:Problem Definition}
\subsection{Preliminaries}\label{Sec:Preliminaries}
Let $\mathcal{G}=(V, E, p)$ be an undirected network, where $V$ is the set of nodes and $E \subseteq V \times V$ is the set of edges.
Let $p:E\rightarrow (0, 1]$ be an assigned probability to indicate the confidence on the existence of an edge.
We consider $p$ particularly for functional networks in this paper, which is why we restrict the range of $p$ to $(0.95, 1]$.
They are usually constructed in a statistical manner with high confidences (more details see Section~\ref{Sec:Experimental settings}) from time series.
If $p(e)=1 ~ \forall e \in E$, then the constructed network is called deterministic.
For the most common symbols used throughout this work, see Table~\ref{Tab:Symbols}.

The sparse network $G^{*}=(V, E^{*})$ after edge sampling is a subgraph of $\mathcal{G}$.
Both have the same number of nodes as we do not consider node aggregation.
To obtain $G^{*}$ in the desired way, we first need to derive scaled degrees, triangles, and wedges associated with nodes.
The basic idea is that each edge in $\mathcal{G}$ contributes to the emergence of observed structural properties, such as degree distribution and community structure.
By scaling down the contribution, one can expect the corresponding properties to be scaled
similarly.
To this end, we include a scaling factor $S \in [0, 1]$ with $\powerps_{e=\{u,v\}}:=p(e)\times S~\forall e \in E$,
which implicitly determines the ratio of preserved edges after sparsification.

\begin{figure}[!htbp]
  \centering
  \includegraphics[width=0.45\textwidth]{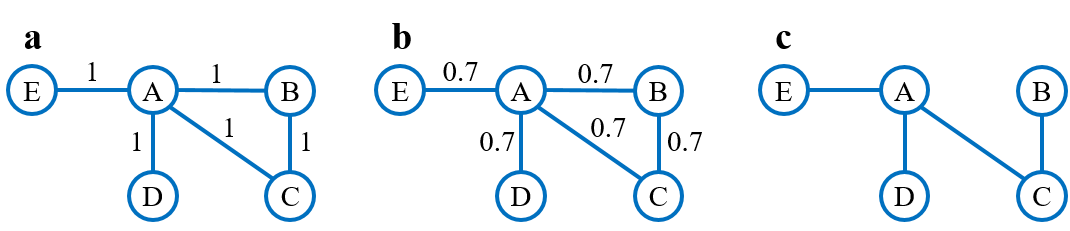}
  \caption{
  {An example of network sparsification via scaled local-property-based edge sampling in this work.
  (a) The original undirected network $\mathcal{G}$ with $\powerps=p=1$ (see Section~\ref{Sec:Preliminaries}) as the confidence on the existence of each edge, indicating that each edge fully contributes to the formation of this network.
  (b) The scaled contribution is assumed to be $\powerps=p \times S=0.7$.
  For example, the expected degree of $A$ and the expected number of triangles that $A$ belongs to become $E[m_{l=2}(A, \mathcal{G})] = \powerps_{A, B} + \powerps_{A, C} + \powerps_{A, D} + \powerps_{A, E} = 2.8$ and $E[m_{l=3}(A, \mathcal{G})] = \powerps_{A, B} \times \powerps_{A, C} \times \powerps_{B, C} = 0.343$ (see Section~\ref{Sec:Sparsification via scaled local properties}), respectively.
  (c) The final sparse network $G^*$ obtained by considering the scaled node properties in (b) as the optimization objective (see Section~\ref{Sec:Sparsification via scaled local properties}) and by using the adapted heuristic method GST$_{2,3}$ (see Section~\ref{Sec:GST}).}
  }
  \label{Fig:Example}
\end{figure}

Since now edges are attached with probabilities, $G^{*}$ should eventually conform well with the scaled structural properties of $\mathcal{G}$ in \emph{expectation}.
That is, we can define the expected degree of a node and the expected number of triangles and wedges the node belongs to, as the scaled local properties to specifically indicate the optimization goal.
We also note that the expected number of wedges is often not as important as the (more prominent) expected number of triangles. 
Therefore, whether to preserve both the expected number of triangles and wedges associated with nodes depends on the application context.

\subsection{Sparsification via scaled local properties}\label{Sec:Sparsification via scaled local properties}
To expand on~\cite{zeng2021selective,zeng2022reductiona} and to take more than local degrees into account, we further consider $3$-size subgraphs and propose a \emph{normalized} definition of network sparsification.
The expected degree of a node and the expected number of triangles and wedges the node belongs to have been defined in the context of uncertain graphs~\cite{parchas2015uncertain} and can also be applied here.

For a given $\mathcal{G}$ and a randomly selected node $u$, all possible neighbors of $u$ form the set $L_{l=2}(u,\mathcal{G})=\{v:\{u,v\}\in {E}\}$ with $|L_{l=2}(u,\mathcal{G})|$ being the degree of $u$ in $\mathcal{G}$.
Given $\powerps$, the corresponding set containing edge contributions is $\powerps_{l=2}(u,\mathcal{G})=\{\powerps_{u,v}:\{u,v\}\in {E}\}$. 
Similarly, all possible triangles that $u$ belongs to form the set $L_{l=3}(u,\mathcal{G})=\{\{u, v, x\}:\{u,v\},\{u,x\},\{v,x\}\in {E}\}$.
The maximum possible number of triangles that $u$ can have in $G^*$ hence is $|L_{l=3}(u,\mathcal{G})|$ and we also have $\powerps_{l=3}(u,\mathcal{G})=\{\{\powerps_{u,v},\powerps_{u,x},\powerps_{v,x}\}:\{u,v\},\{u,x\},\{v,x\}\in {E}\}$.

By assuming the independence of edge probabilities~\cite{parchas2014pursuit}, the expected degree of $u$ as well as the expected number of triangles and wedges that $u$ belongs to (respectively), are derived using the linearity of expectation as~\cite{parchas2015uncertain}: 
\begin{equation}\label{Eq_Expected_degree}
  E[m_{l=2}(u, \mathcal{G})] := \sum_{i=1}^{|\powerps_{l=2}(u,\mathcal{G})|} \powerps_{l=2}(u,\mathcal{G})_{i}
\end{equation}
\begin{equation}\label{Eq_Expected_triangle}
  E[m_{l=3}(u, \mathcal{G})] := \sum_{i=1}^{|\powerps_{l=3}(u,\mathcal{G})|} \prod_{j=1}^{3} \powerps_{l=3}(u,\mathcal{G})_{j}
\end{equation}
\begin{equation}\label{Eq_Expected_wedge}
  \begin{split}
    E[m_{l=w}(u, \mathcal{G})] & := \frac{1}{2} \left(E[m_{l=2}(u, \mathcal{G})^2] - E[m_{l=2}(u, \mathcal{G})]\right) \\
    & - E[m_{l=3}(u, \mathcal{G})]
  \end{split}
\end{equation}
where $i$ and $j$ iterate over the members of the sets $\powerps_{l=2}(u,\mathcal{G})$ and $\powerps_{l=3}(u,\mathcal{G})$, respectively.
An example for the calculation of Eqs.~(\ref{Eq_Expected_degree}) and~(\ref{Eq_Expected_triangle}) is given in Figure~\ref{Fig:Example}b.
For Eq.~(\ref{Eq_Expected_wedge}), if we let $X$ be a discrete random variable with $X=m_{l=2}(u, \mathcal{G})=\{x_i|x_i\in[0, |L_{l=2}(u,\mathcal{G})|]\}$, then it represents all possible degrees of $u$ during the edge sampling process.
To obtain $E[X^2]$, we calculate $Pr(X=x_i)$ by using dynamic programming based on $\powerps_{l=2}(u,\mathcal{G})$~\cite{bonchi2014core}.

In a sparse subgraph $G^{*}$, each node should be as close as possible to its expected basic local properties.
For this, we define the \emph{normalized} distance from the current degree ($l=2$) of $u$ and the current number of triangles ($l=3$) and wedges ($l=w$) that $u$ belongs to in $G'$, to their expectations:
\begin{equation}\label{Eq_Distance}
  \Delta m_{l}(u, G') := \frac{1}{|L_{l}(u, \mathcal{G})|}|m_{l}(u, G') - E[m_{l}(u, \mathcal{G})]|
\end{equation}
where $\frac{1}{|L_{l}(u, \mathcal{G})|}$ is a normalization factor to distinguish the positions of different nodes.
It emphasizes that the sparsification by edge sampling is built on top of the original network.
Note that previous studies~\cite{zeng2021selective,zeng2022reductiona,parchas2014pursuit,parchas2015uncertain} ignore this factor.
We demonstrate its importance in Section~\ref{Sec:Complex property preservation}.
The total distance for a subgraph $G'$ is therefore defined as:

\begin{defn}
  Given an undirected network $\mathcal{G}=(V, E, p)$ and scaled local properties (on the expected degree of each node and the expected number of triangles and wedges each node belongs to) to be preserved, the distance of any subgraph $G' \subseteq \mathcal{G}$ to its overall expectation is:
  \begin{equation}\label{Eq_Total_Distance}
    \Delta m_{l=2,3,w}(G') :=  \sum_{u\in V}\sum_{l=2,3,w}\Delta m_{l}(u, G')
  \end{equation}
\end{defn}

The network sparsification problem via edge sampling is therefore defined as:
\newcommand{\argmin}{\operatornamewithlimits{argmin}}
\begin{defn}
  (Sparsification via scaled local properties). Given an undirected network $\mathcal{G}=(V, E, p)$, find a subgraph $G^{*}=(V, E^{*})$ such that:
  \begin{equation}\label{Eq_Sparsification_By_Local_properties}
    G^{*} := \argmin_{G' \subseteq \mathcal{G}} \Delta m_{l=2,3,w}(G')
  \end{equation}
\end{defn}
By default, this is meant as the $\argmin$ for all three properties together.
In our experiments, we will also look at subsets thereof (degrees and triangles), though.
According to Ref.~\cite{parchas2014pursuit}, for $l=2$ this problem is a special case of the closest vector problem, which is $\mathcal{NP}$-hard~\cite{micciancio2001hardness}.
As our problem is a generalization, it is $\mathcal{NP}$-hard, too.
We hence aim at providing heuristic solutions that are fast and accurate enough for practical purposes.

\section{The game-theoretic sparsification with tolerance (GST)}\label{Sec:GST}

\citet{parchas2015uncertain} proposed a game-theoretic framework for uncertain graph sampling.
It consists of an exact potential game with convergence to a Nash equilibrium based on best-response dynamics~\cite{monderer1996potential}.
We adapt this framework to sparsification and include a tolerance factor that allows to
terminate when the progress is below a user-specified threshold.

The basic idea is that each edge $e=\{x, y\}\in {E}$ in a given $\mathcal{G}$ is modeled as a player involved in an exact potential game.
In this game, the gain change in the individual cost function is reflected in a global potential function.
Specifically, each edge decides whether to be preserved (binary states: $1$ for preservation) in the final sparse graph $G^{*}$.
Suppose that the decision of $e$ changes the current $G'=(V, E')$ into $G''=(V, E'')$;
the current state of $e$ is updated only if this leads to a positive \emph{gain} change ($g(e)>0$), which is defined as~\cite{parchas2015uncertain}:
\begin{equation}\label{Eq_Gain_Global}
    g(e)  :=  \sum_{u\in V}\sum_{l=2,3,w}(\Delta m_{l}(u, G') - \Delta m_{l}(u, G'')),
\end{equation}
where the sum of $\Delta m_{l}(u, G')$ is the gain (or the distance to the overall expectation) of $e$ by retaining its current state, while the sum of $\Delta m_{l}(u, G'')$ is the gain of $e$ by changing its state.
The result is the overall gain change resulting from the global potential function where each node $u \in V$ is involved.
Recall that we consider basic local properties: degrees and $3$-node subgraphs (wedges and triangles).
Only a limited number of nodes are therefore affected by a change in $e$ in Eq.~(\ref{Eq_Gain_Global}).
These nodes include $x$, $y$, and common neighbors of $x$ and $y$ in $G'$, forming a set we denote as $L(e)$.
Therefore, as proved in~\cite{parchas2015uncertain}, Eq.~(\ref{Eq_Gain_Global}) is equivalent to:
\begin{equation}\label{Eq_Gain_Individual}
    g(e) := \sum_{v\in L(e)}\sum_{l=2,3,w}(\Delta m_{l}(v, G') - \Delta m_{l}(v, G''))
\end{equation}
which represents the change of the individual cost function.
The equivalence of Eqs.~(\ref{Eq_Gain_Global}) and~(\ref{Eq_Gain_Individual}) ensures that this game is an exact potential game.
The best-response dynamics -- that each edge repeatedly changes its state based on the decisions of all others -- on the exact potential game guarantees the convergence to a Nash equilibrium~\cite{monderer1996potential}.
That is, if the corresponding algorithm models this process, it will terminate.

Algorithm~\ref{Alg:GST} presents the pseudocode of GST, which models such an exact potential game.
We emphasize again that although we consider by default the preservation of the expected number of wedges ($l=w$) in GST, empirical studies still need to compare two cases: with and without $l=w$.
The inputs include an undirected network $\mathcal{G}$ and two important values, the tolerance $T$ for early termination and the scaling factor $S$ for sparsification.
Stage I (lines 1-5) computes the expected local properties based on $\mathcal{G}$ and $S$.
The computation of Eqs.~(\ref{Eq_Expected_degree}) and~(\ref{Eq_Expected_triangle}) are parallelized since they need only local information. 
Stage II initializes the current subgraph $G'$ with the entire set $E$ in line 6.
The $m_{l}(u, G')$ in line~8 are therefore exactly the same as $|L_{l}(u, \mathcal{G})|$ for $l=2,3,w$.
$L_{new}$ represents the set of all affected nodes and is initialized with the entire set $V$.
We include another array $Gain[|V|]$ for recording $\Delta m_{l=2,3,w}(G')$ during iterations.
Starting from line 10, the algorithm proceeds in rounds.
In each round, given an edge $e$ incident to a node in $L$, it first finds all affected nodes in $G'$ by the decision of $e$.
That is, $L(e)$ includes $x$, $y$, and common neighbors of $x$ and $y$ in $G'$.
Then, it computes $g(e)$ induced by assuming that $e$ changes its current state.
If $e\in E'$ and the removal of $e$ leads to a positive $g(e)$, then $e$ changes from $1$ to $0$.
If $e\notin E'$ and the preservation of $e$ gives a positive $g(e)$, then $e$ switches from $0$ to $1$. The iteration stops based on the progress in the gain in relation to the threshold $T$.

\SetKwHangingKw{KwHData}{Data$\rightarrow$}
\SetKwInput{KwIn}{Input}%
\SetKwInput{KwOut}{Output}%
\SetKwInput{KwData}{Data}%
\SetKwInput{KwResult}{Result}%
\SetKw{KwTo}{to}
\SetKw{KwRet}{return}%
\SetKw{Return}{return}%
\SetKwBlock{Begin}{begin}{end}%
\SetKwRepeat{Repeat}{repeat}{until}%
\SetKwIF{If}{ElseIf}{Else}{if}{then}{else if}{else}{end if}%
\SetKwSwitch{Switch}{Case}{Other}{switch}{do}{case}{otherwise}{end case}{end switch}%
\SetKwFor{For}{for}{do}{end for}%
\SetKwFor{ForPar}{for}{do in parallel}{end forpar}
\SetKwFor{ForEach}{for each}{do}{end foreach}%
\SetKwFor{ForAll}{for all}{do}{end forall}%
\SetKwFor{While}{while}{do}{end while}%
\begin{algorithm}[!t]
  \footnotesize
  \DontPrintSemicolon
  \KwIn{An undirected network $\mathcal{G}=(V, E, p)$, tolerance factor $T=0.01$, scaling factor $S\in [0, 1]$}
  \KwOut{$G^{*}=(V, E^{*})$}

  \tcp*{Stage I (The expected basic properties)}
  $\powerps \gets p \times S$ \\
  \ForPar{$u \in V$}
  {
    Compute Eqs.~(\ref{Eq_Expected_degree}) and (\ref{Eq_Expected_triangle}), $|L_{l=2}(u,\mathcal{G})|$, and $|L_{l=3}(u,\mathcal{G})|$
  }
  \ForEach{$u \in V$} 
  { 
    Compute Eq.~(\ref{Eq_Expected_wedge}) and $|L_{l=w}(u,\mathcal{G})|$}

  \tcp*{Stage II (Sparsification)}
  $E' \gets E$\\
  \ForPar{$u \in V$} 
  { 
    $m_{l}(u, G') \gets |L_{l}(u, \mathcal{G})|$ ($l=2,3,w$, respectively)
  }
  $L_{new} \gets V$; {$Gain[|V|] \gets 0$}; $r \gets 0$ \\
  \Repeat{$r \geq 2$ and $Gain[r-1]-Gain[r] \leq T$}
  {
    $L \gets L_{new}$; $L_{new} \gets \emptyset$ \\
    \ForEach{$e=\{x,y\} \in E$ incident (in $\mathcal{G}$) to a node in $L$}
    {
      $L(e) \gets \{x, y\}\cup\{u \in V: \{u,x\} \in E' \wedge \{u,y\} \in E'\}$ \\
      Compute $g(e)$ by Eq.~(\ref{Eq_Gain_Individual}) \\
      \If{$g(e) > 0$}
      {
        \If{$e \in E'$} 
        {
          $E' \gets E'\setminus \{e\}$; $L_{new} \gets L_{new} \cup L(e)$
        }
        \Else{$E' \gets E'\cup \{e\}$; $L_{new} \gets L_{new} \cup L(e)$}
      }
    }
    $r \gets r + 1$; $Gain[r] \gets \Delta m_{l=2,3,w}(G')$ \\
  }
  \Return{$E^{*} \gets E'$}
  \caption{Game-theoretic sparsification with tolerance (GST)}\label{Alg:GST}
\end{algorithm}

Regarding (sequential) time complexity, we note for Stage I that computing Eq.~(\ref{Eq_Expected_degree}) takes $\mathcal{O}(|E|)$ time.
When computing triangles, we use a merge-based intersection operation between $u$ and each of its neighbors, since each node already has a sorted neighbor set.
Computing Eq.~(\ref{Eq_Expected_triangle}) therefore takes $\mathcal{O}(d_{max}|E|)$ time in total, where $d_{max} = \max\{|L_{l=2}(u, \mathcal{G})|:u\in V\}$ is the maximum (possible) degree in $\mathcal{G}$.
According to~\cite{ortmann2013triangle}, an even tighter bound is $\mathcal{O}(a(\mathcal{G})|E|)$, with $a(\mathcal{G})$ being the arboricity of $\mathcal{G}$.
Computing Eq.~(\ref{Eq_Expected_wedge}) via dynamic programming also takes $\mathcal{O}(d_{max}|E|)$ time.
Hence, the total time complexity of Stage I is $\mathcal{O}(d_{max}|E|)$.

Stage II depends mostly on the time spent on the repeat-loop.
Finding $L(e)$ involves a linear-time intersection operation with $\mathcal{O}(d_{max})$ time each for two already sorted neighbor sets.
For similar reasons as in Stage I, the for-loop takes $\mathcal{O}(d_{max}|E|)$ time
(per iteration of the repeat-loop).
Stage II therefore needs $\mathcal{O}(rd_{max}|E|)$ in total, where $r$ is the number of
iterations of the repeat-loop.
Thus, in total, the time complexity of Algorithm~\ref{Alg:GST} is $\mathcal{O}(rd_{max}|E|)$.
We show in Section~\ref{Sec:Basic property preservation} how the tolerance threshold $T$ affects the convergence positively.
Moreover, we present in Section~\ref{Sec:Running time consumption} the empirical running times of both stages.

\section{Applications to climate data}\label{Sec:Applications to climate data}
We assess the performance of GST by answering the following three questions in Secs.~\ref{Sec:Basic property preservation}, ~\ref{Sec:Complex property preservation}, and~\ref{Sec:Running time consumption}, respectively:
\begin{itemize}
  \item[\textbf{Q1:}] How well does GST generate a sparse $G^*$ preserving scaled local properties?
  \item[\textbf{Q2:}] How well does GST generate a sparse $G^*$ preserving non-local / complex properties?
  \item[\textbf{Q3:}] How is the running time of GST?
\end{itemize}

\setlength\extrarowheight{3pt}
\begin{table}[b]
\centering
  \footnotesize
  \caption{Characteristics of data sets}\label{Tab:Characteristics of data sets}
  \setlength\arrayrulewidth{1pt}
  \setlength{\tabcolsep}{3pt}
  \begin{tabularx}{0.5\textwidth}{cccccc}
    \hline
    Network       & Nodes ($|V|$) & edges ($|E|$) & $\frac{|E|}{|V|}$ & Edge confidence ($p$) \\
    \hline
    Glo\_ERA5SP   & 7,320         & 593,736       & 81.11             & 1                     \\
    Glo\_ERA5ST   & 7,320         & 882,102       & 120.51            & 1                     \\
    Glo\_TRMM & 36,000        & 2,139,214     & 59.42             & [0.99, 1]             \\
    ASM\_TRMM & 20,000        & 1,771,609     & 88.58             & [0.99, 1]             \\
    \hline
  \end{tabularx}
\end{table}

\subsection{Experimental settings}\label{Sec:Experimental settings}
\textbf{(1) Climate data sets.}
Our particular focus on climate data is mainly driven by studies of complex climate phenomena using complex networks during the last two decades.
The reconstructed functional climate networks can be large especially when a high spatial resolution is considered.
Four functional networks are summarized in Table~\ref{Tab:Characteristics of data sets}.
If $p=1$, then the corresponding networks are completely deterministic:
\begin{itemize}
  \item Glo\_ERA5SP: we use the time series of daily surface level pressure (SP) within the June-July-August season from 1998 to 2019 from ERA5 reanalysis data~\cite{hersbach2020era5b}, with the global spatial resolution of 1$^{\circ}$ $\times$ 1$^{\circ}$.
        The functional network reconstruction process is adapted from~\cite{gupta2021complex} by viewing grid points as nodes and using Spearman correlation as the similarity between time series.
  \item Glo\_ERA5ST: it is the same as Glo\_ERA5SP, but using daily surface level temperature (ST)~\cite{hersbach2020era5b}.
  \item Glo\_TRMM: we consider the observational data of global precipitation from Tropical Rainfall Measuring Mission 3B42v6 product (TRMM)~\cite{huffman2007trmmb}.
  Time series represent the daily rainfall sums within the June-July-August season from 1998 to 2019 with a spatial resolution of $1^{\circ} \times 1^{\circ}$.
  As precipitation data are spiking series, we adopt event synchronization (ES) as a nonlinear similarity measure~\cite{quianquiroga2002eventb}.
        By treating grid points as nodes, the reconstruction process is the same as  in~\cite{su2022climatic}.
  \item ASM\_TRMM: it is the same as Glo\_TRMM, but focuses on a relatively small region, i.e., 
  the Asian summer monsoon (ASM) region, instead of the global scale.
\end{itemize}

\textbf{(2) Baselines.}
We compare GST with four baselines.
\citet{zeng2021selective,zeng2022reductiona} studied preserving the expected degree of each node and adapted two approximate methods similar to those in uncertain graph sampling~\cite{parchas2015uncertain}.
\citet{parchas2015uncertain}  concluded that among all approximate methods they proposed, the game-theoretic framework generates better representative instances.
We directly adapt this framework for network sparsification.
In particular, GST$_{2}$ preserves only the scaled degrees and corresponds to~\cite{zeng2021selective,zeng2022reductiona}.
Therefore, the first comparison is between GST$_{2}$ and our extensions GST$_{2,3}$/GST$_{2,3,w}$, where $3$ and $w$ denote 3-node subgraphs (triangles and wedges, respectively) associated with each node (see Section~\ref{Sec:Basic property preservation}).
Other three well-known sampling methods are \emph{local degree} (LD)~\cite{hamann2016structurepreserving}, \emph{local jaccard similarity} (LJS)~\cite{satuluri2011local}, and \emph{random edge} (RE)~\cite{sadhanala2016graph} (see Section~\ref{Sec:Complex property preservation}).
We choose them due to their effectiveness in preserving the overall connectivity (by LD), community structure (by LJS), and spectral property (by RE), at least in non-functional networks.
They have been systematically compared in~\cite{hamann2016structurepreserving} and implemented in \textsc{NetworKit}~\cite{staudt2016networkita}, a tool suite for scalable network analysis.

\textbf{(3) Evaluation metrics.}
For \textbf{Q1}, we analyze the extent to which the expected degree and the expected number of 3-node subgraphs associated with each node are preserved, even when bearing some loss with the inclusion of a tolerance threshold $T$ in GST.
The four measures below are used (see Section~\ref{Sec:Basic property preservation}):
\begin{itemize}
    \item \emph{Node-wise distance distribution}: $\Delta_{2,3,w}(G^{*}) = \Delta m_{l=2}(u, G^{*}) + \Delta m_{l=3}(u, G^{*}) + \Delta m_{l=w}(u, G^{*})$.
    It represents the summarized overall distance of the local properties (degrees, triangles, and wedges) for each node in the generated $G^{*}$ to the expectation.
    Hence, $\Delta_{2,3,w}(G^{*})$ is a sequence of length $|V|$.
    The minimization of the sum of $\Delta_{2,3,w}(G^{*})$ over all nodes corresponds to the objective of Eq.~(\ref{Eq_Sparsification_By_Local_properties}).
    \item \emph{Mean distance}: $\overline{\Delta}_{2,3,w}(G^{*}) = \frac{1}{|V|} \sum_1^{|V|} \Delta_{2,3,w}(G^{*})$.
    \item \emph{Convergence of mean distance}: $\overline{\Delta}_{2,3,w}(G') = \frac{1}{|V|}(\Delta m_{l=2}(u, G') + \Delta m_{l=3}(u, G') + \Delta m_{l=w}(u, G'))$.
    This measure is designed for convergence analysis, since it is based on the current $G'$ (at line 20 of Algorithm~\ref{Alg:GST}) instead of $G^{*}$.
    \item \emph{Cumulative time}: total time spent until the current iteration (lines 10-25 in Algorithm~\ref{Alg:GST}), also for empirical convergence analysis.
\end{itemize}

For \textbf{Q2}, the following property queries are considered:
\emph{macroscopic}: the global clustering coefficient and largest connected component;
\emph{mesoscopic}: community structure and betweenness centrality;
\emph{microscopic}: degree and local clustering coefficient.
Both mesoscopic and microscopic queries have been applied in functional climate network analysis~\cite{su2022climatic,gupta2021complex}.
In particular, computing the exact betweenness values is in practice very expensive for the unsparsified network.
Therefore, we use \emph{ApproxBetweenness} from \textsc{NetworKit}~\cite{staudt2016networkita} with a guarantee that the error is no larger than 0.01, with a probability of at least 0.9.
Measures used to estimate the similarity between the properties calculated from $G^*$ and $\mathcal{G}$ are (see Section~\ref{Sec:Complex property preservation}):
\begin{itemize}
    \item Average \emph{Deviation}~\cite{hamann2016structurepreserving}: we analyze the deviation of the macroscopic properties in $G^*$ from those in $\mathcal{G}$, because these properties are single-valued representations.
    \item Average \emph{Adjusted rand index} (ARI)~\cite{vinh2010information}: this one is particularly used for giving the similarity between two clusterings obtained based on the final sparse network $G^*$ and the original network $\mathcal{G}$, respectively.
    \item Average \emph{Spearman} rank correlation coefficient~\cite{hamann2016structurepreserving}: microscopic properties are node-wise representations, therefore similarities are estimated using correlation {with a significance level of $P < 0.01$}.
\end{itemize}

The estimation process is as follows.
Taking the comparison between GST$_{2,3}$($T=0.01$) and LD as an example, we first generate 100 sparse networks $G^*$ for a given $\mathcal{G}$, based on GST$_{2,3}$($T=0.01$).
Then, another 100 sparse networks, say LD$_{G^*}$, are created by using LD with the preservation ratio of edges calculated based on the edge ratio between $G^*$ and $\mathcal{G}$.
Assuming the query on community structure, we apply the parallel Louvain method~\cite{staudt2016engineeringa} from \textsc{NetworKit}~\cite{staudt2016networkita} to $\mathcal{G}$, $G^*$, and LD$_{G^*}$, respectively.
We then compute the ARI between the highest-quality (out of 100 repeated runs) community structures obtained from $\mathcal{G}$ and each $G^*$; the same process is applied to $\mathcal{G}$ and each LD$_{G^*}$.
One can notice that it is hard to ensure that one edge sampling method outperforms the rest for all of these property queries.
Therefore, we need additional measures summarizing all queries, instead of checking them one by one (see Section~\ref{Sec:Complex property preservation}):
\begin{itemize}
    \item \emph{Ranking distribution}: for each given scaling factor $S$, each query task gives a ranking between GST, LD, LJS and RE, from 1 to 4.
    We summarize all rankings of each method for different $S$ and property queries.
    \item \emph{Mean ranking}: the mean of all rankings of each method.
\end{itemize}

For \textbf{Q3}, to give a fair comparison between GST, LD, LJS, and RE (see Section~\ref{Sec:Running time consumption}), we choose a single-threaded environment
without parallelization. The comparison shows the average running time over 100 runs.

\subsection{Basic property preservation}\label{Sec:Basic property preservation}
\begin{figure}[!htbp]
  \centering
  \includegraphics[width=0.5\textwidth]{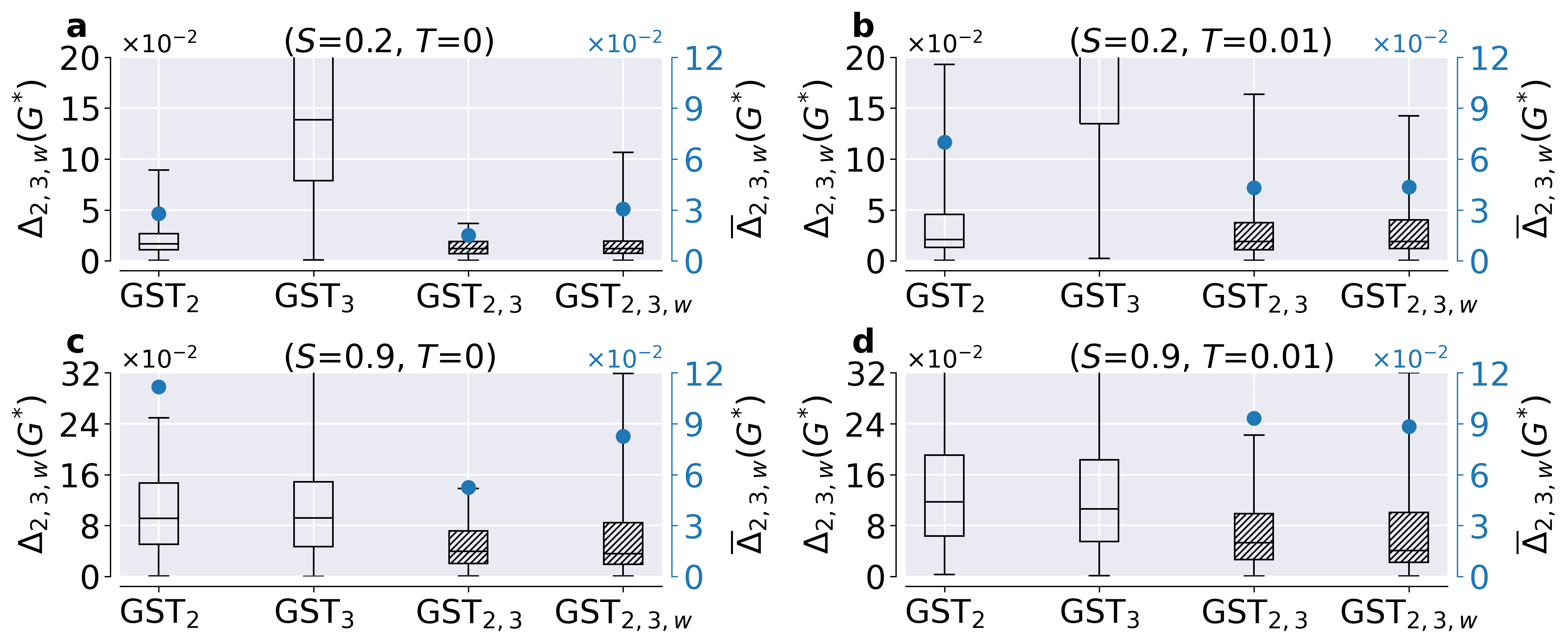}
  \caption{The distribution (left y-axis) and mean (right y-axis) of $\Delta_{2,3,w}(G^{*})$ based on $G^*$, versus different preservation scenarios for Glo\_ERA5SP.
  Boxplots show how close 0\%, 25\%, 50\%, 75\% and 95\% nodes are to their expected local properties (2, 3, and w for degrees, triangles, and wedges, respectively).
  The suffixes of GST represent the local properties chosen to be preserved.
  (a) GST($S$=0.2, $T=0$). (b) GST($S$=0.2, $T=0.01$). (c) GST($S$=0.9, $T=0$). (d) GST($S$=0.9, $T=0.01$).
  (a) and (b) produce a sparser structure due to a smaller $S$.
  This figure indicates the benefit of preserving both the expected degree of each node \emph{and} the expected number of 3-node subgraphs each node belongs to.}
  \label{Fig:ERA5_SP_2_3_23_23w_and_T}
\end{figure}

\begin{figure}[!htbp]
  \centering
  \includegraphics[width=0.5\textwidth]{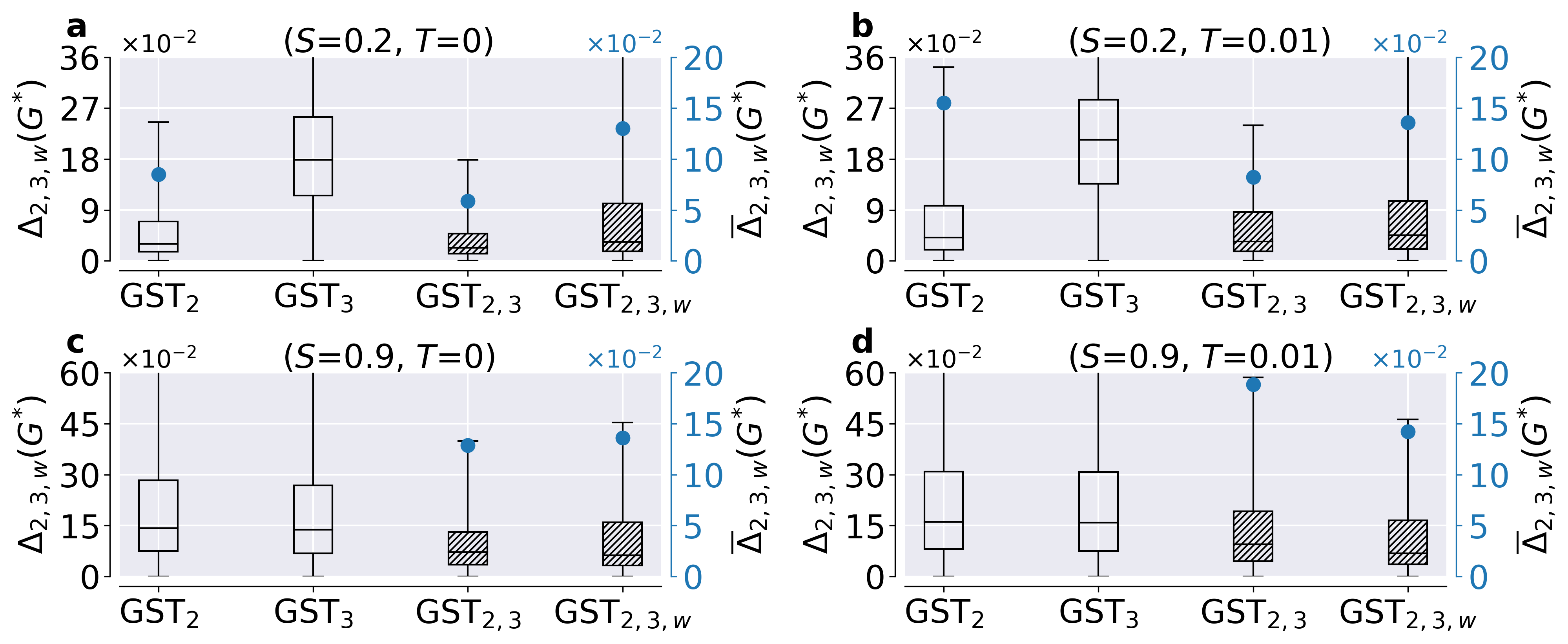}
  \caption{Same as Figure~\ref{Fig:ERA5_SP_2_3_23_23w_and_T} but for Glo\_ERA5ST.}
  \label{Fig:ERA5_ST_2_3_23_23w_and_T}
\end{figure}

\begin{figure}[!htbp]
  \centering
  \includegraphics[width=0.5\textwidth]{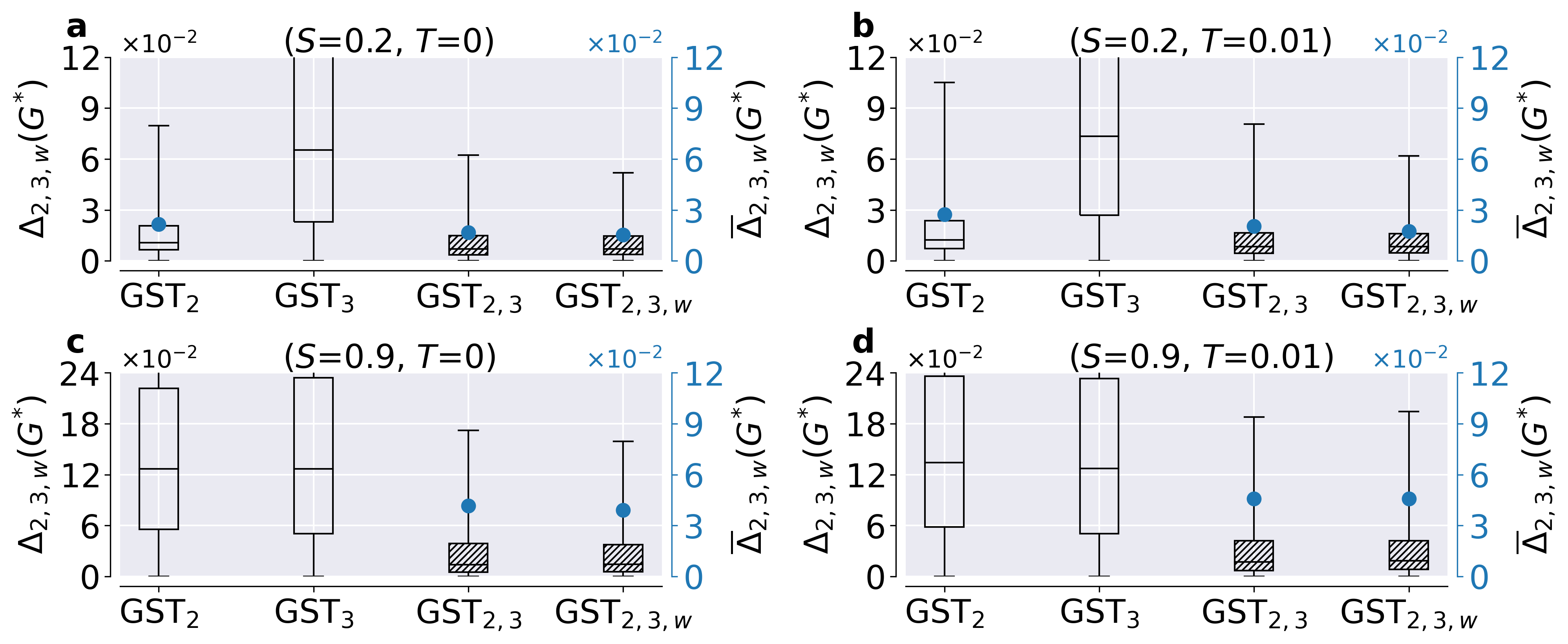}
  \caption{Same as Figure~\ref{Fig:ERA5_SP_2_3_23_23w_and_T} but for Glo\_TRMM.}
  \label{Fig:TRMM_2_3_23_23w_and_T}
\end{figure}

\begin{figure}[!htbp]
  \centering
  \includegraphics[width=0.5\textwidth]{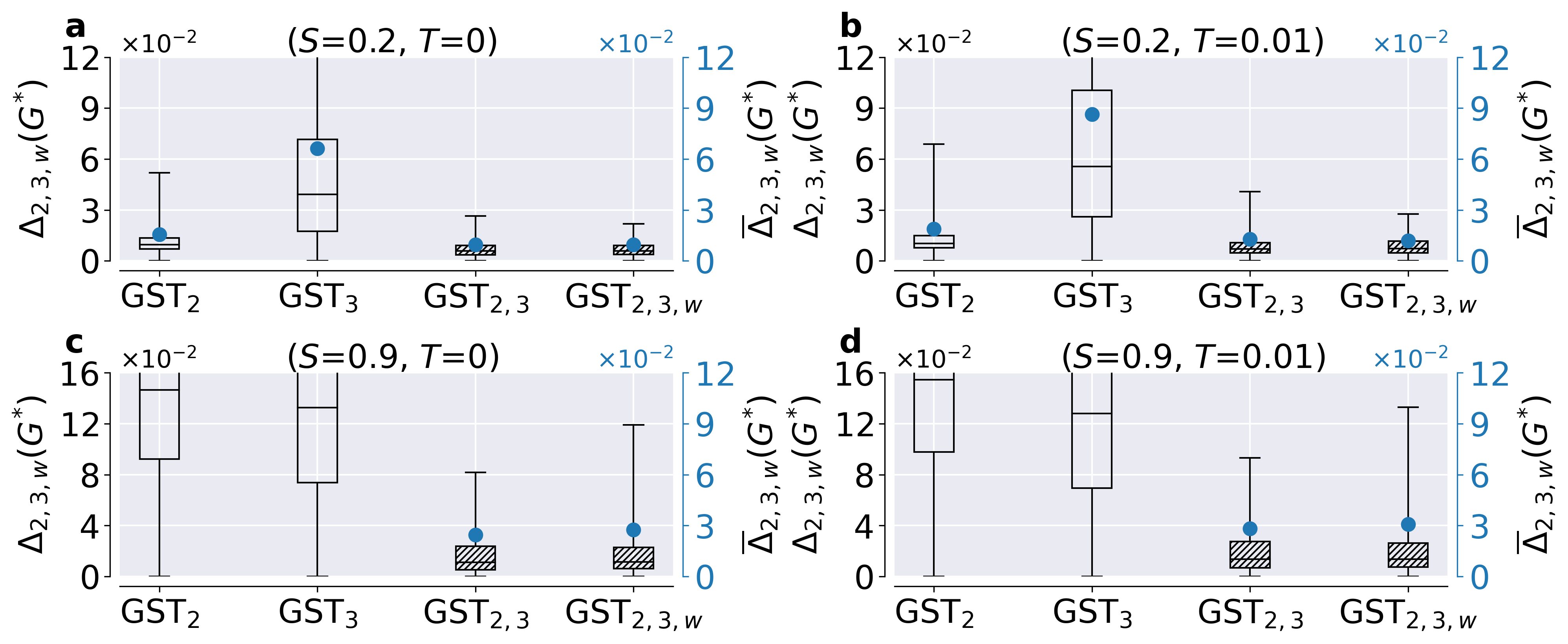}
  \caption{Same as Figure~\ref{Fig:ERA5_SP_2_3_23_23w_and_T} but for ASM\_TRMM.}
  \label{Fig:ASM_2_3_23_23w_and_T}
\end{figure}

We show the distribution of $\Delta_{2,3,w}(G^{*})$ using boxplots and $\overline{\Delta}_{2,3,w}(G^{*})$ in Figures~\ref{Fig:ERA5_SP_2_3_23_23w_and_T},~\ref{Fig:ERA5_ST_2_3_23_23w_and_T},~\ref{Fig:TRMM_2_3_23_23w_and_T} and~\ref{Fig:ASM_2_3_23_23w_and_T}.
The preservation scenarios which include 3-node subgraphs (triangles and wedges) are highlighted with hatches (see GST$_{2,3}$ and GST$_{2,3,w}$).
Regarding \textbf{Q1}, we conclude that preserving both the expected degrees \emph{and} the expected number of 3-node subgraphs generates sparse structures closer to the expectation than only considering degrees.
This fact holds even when the tolerance is set to $T>0$.
From here on we focus only on GST$_{2,3}$ and GST$_{2,3,w}$.


\begin{figure}[!htbp]
  \centering
  \includegraphics[width=0.5\textwidth]{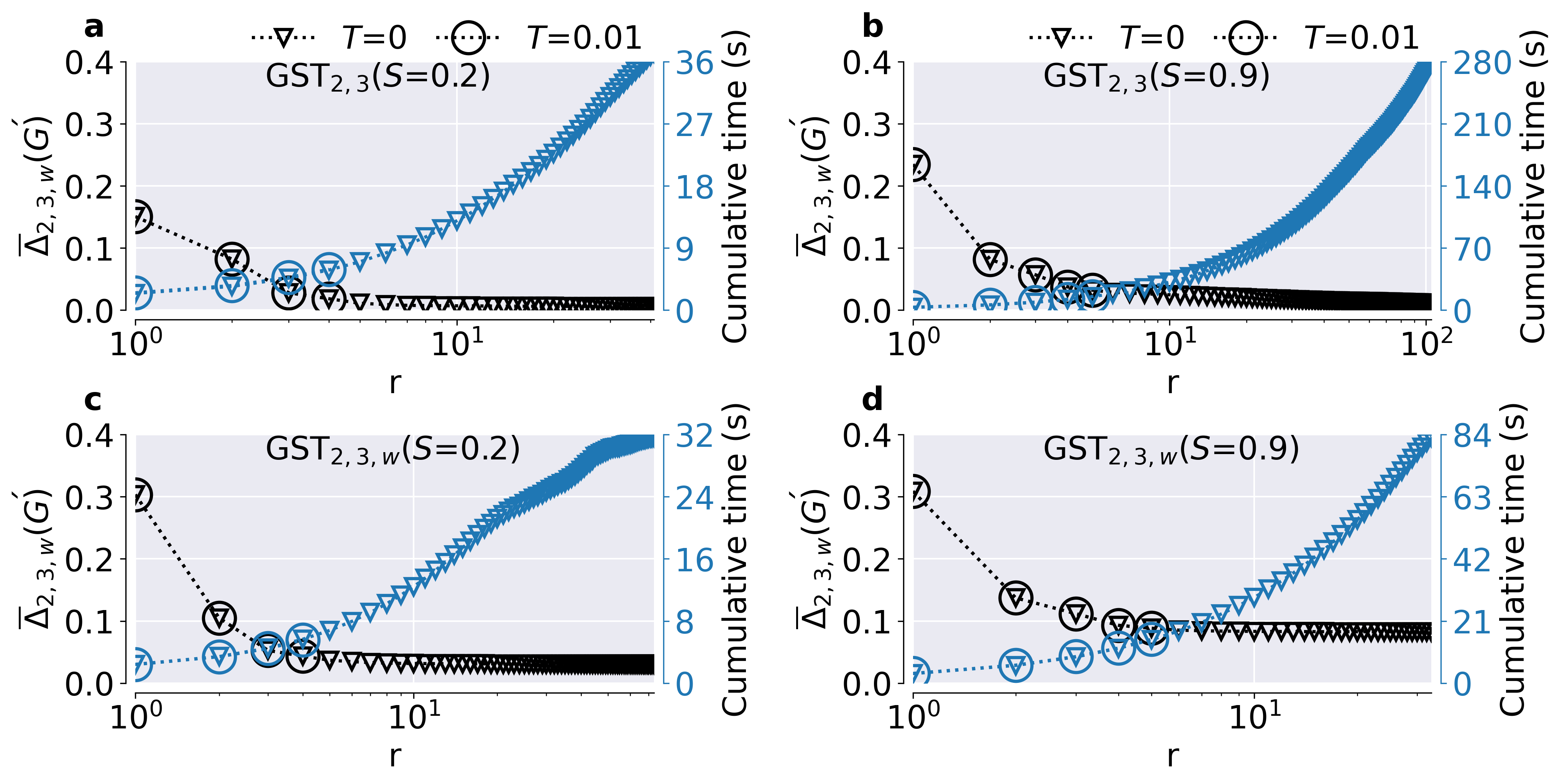}
  \caption{The convergence ($\overline{\Delta}_{2,3,w}(G')$) and cumulative time of GST base on the current $G'$, versus the number of iterations $r$ for Glo\_ERA5SP. (a) GST$_{2,3}$($S$=0.2). (b) GST$_{2,3}$($S$=0.9). (c) GST$_{2,3,w}$($S$=0.2). (d) GST$_{2,3,w}$($S$=0.9).
  Only GST$_{2,3}$ and GST$_{2,3,w}$ are given here since Figures~\ref{Fig:ERA5_SP_2_3_23_23w_and_T},~\ref{Fig:ERA5_ST_2_3_23_23w_and_T},~\ref{Fig:TRMM_2_3_23_23w_and_T} and~\ref{Fig:ASM_2_3_23_23w_and_T} confirme the better performance when 3-node subgraphs (triangles and wedges) are considered for preservation.
  This figure indicates the inclusion of the tolerance factor $T=0.01$ (blue circles) facilitates (at least 4 times faster) the convergence of GST while guaranteeing the quality of the final sparse structure close to $T=0$ (black circles).
  }
  \label{Fig:ERA5_SP_23_23w_and_T}
\end{figure}
\begin{figure}[!htbp]
  \centering
  \includegraphics[width=0.45\textwidth]{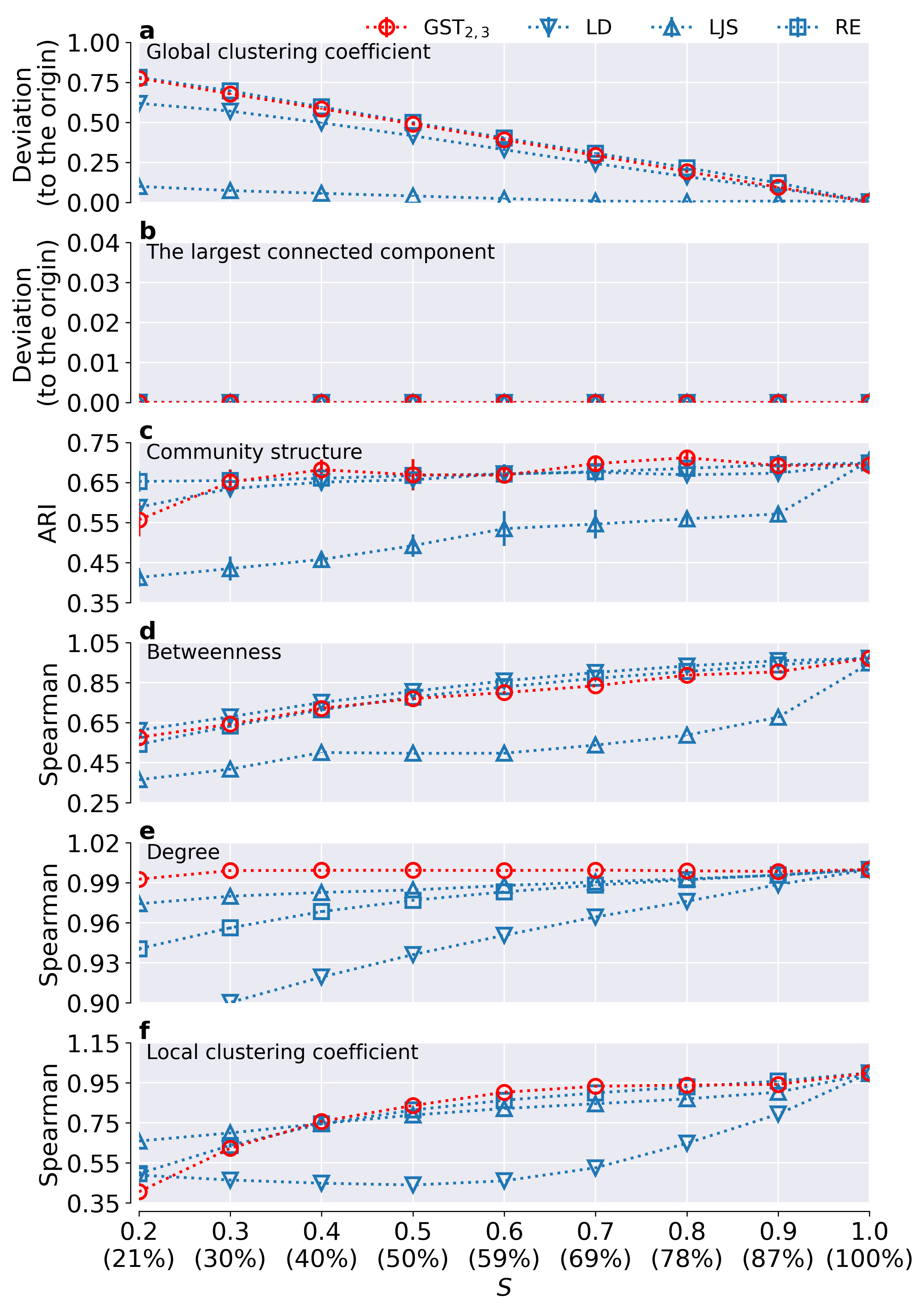}
  \caption{Comparisons between GST$_{2,3}$($T=0.01$), LD, LJS and RE on six structural queries for Glo\_ERA5SP.
  Each scaling factor on the x-axis is attached with the exact ratio of preserved edges in brackets.
  The GST$_{2,3}$($T=0.01$) is highlighted in red.
  This figure indicates that there is no single method that performs better for all of these queries.}
  \label{Fig:ERA5_SP_23}
\end{figure}

\begin{figure}[!htbp]
  \centering
  \includegraphics[width=0.5\textwidth]{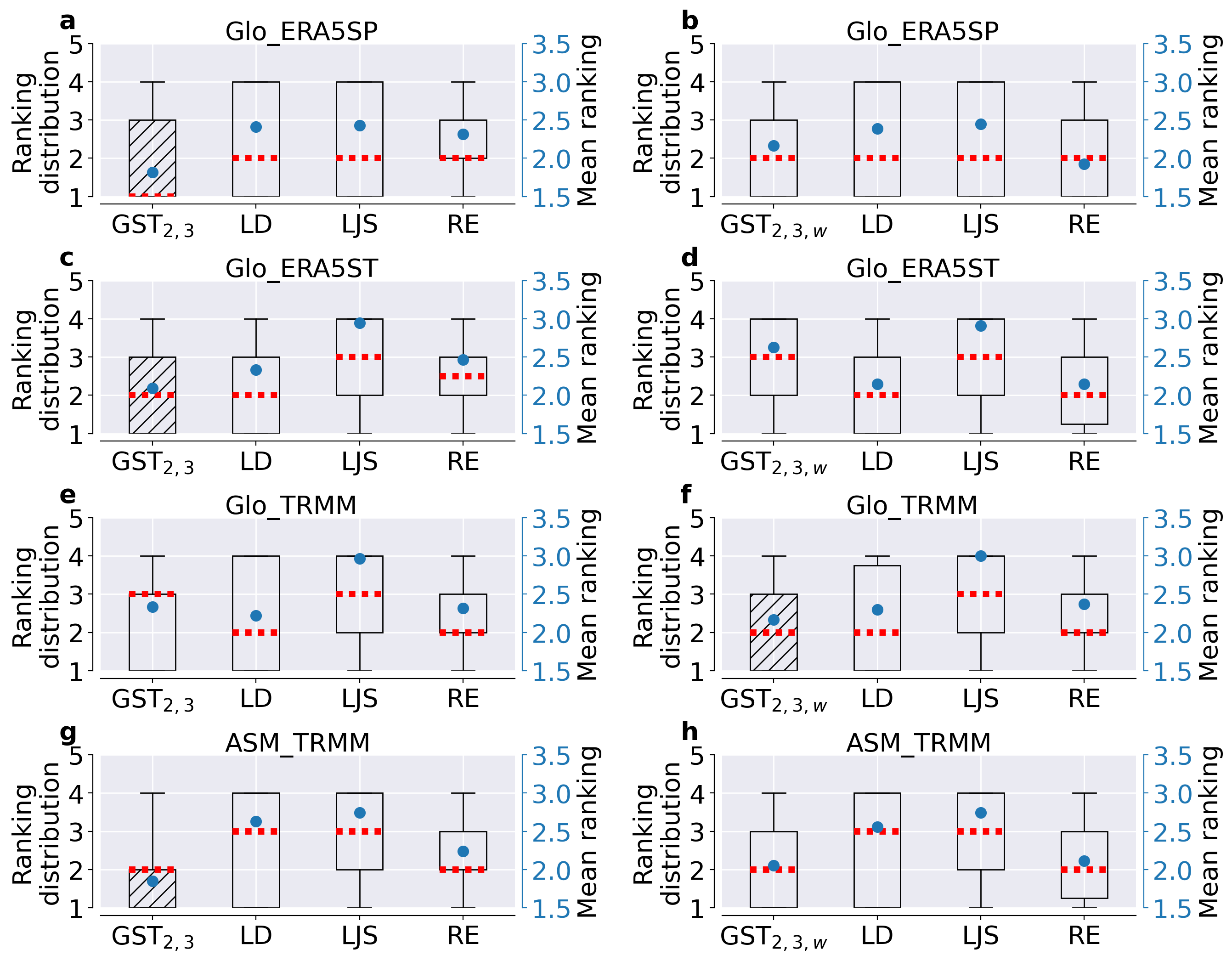}
  \caption{The ranking distribution and mean ranking of GST (GST$_{2,3}$($T=0.01$) and GST$_{2,3,w}$($T=0.01$)), LD, LJS and RE, summarized over six property queries for four networks.
  (a) and (b) Glo\_ERA5SP. (a) is also the summarized rankings of Figure~\ref{Fig:ERA5_SP_23}.  (c) and (d) Glo\_ERA5SP. (e) and (f) Glo\_TRMM. (g) and (h) ASM\_TRMM.
  For each network, the best sampling method is highlighted with hatches and the red dash line is the median of the ranking distribution.
  This figure indicates that the overall performance of GST by preserving both scaled degrees and 3-node subgraphs is better than filtering-based approaches LD, LJS, and RE.}
  \label{Fig:Ranking_all}
\end{figure}
\begin{figure}[!htbp]
  \centering
  \includegraphics[width=0.3\textwidth]{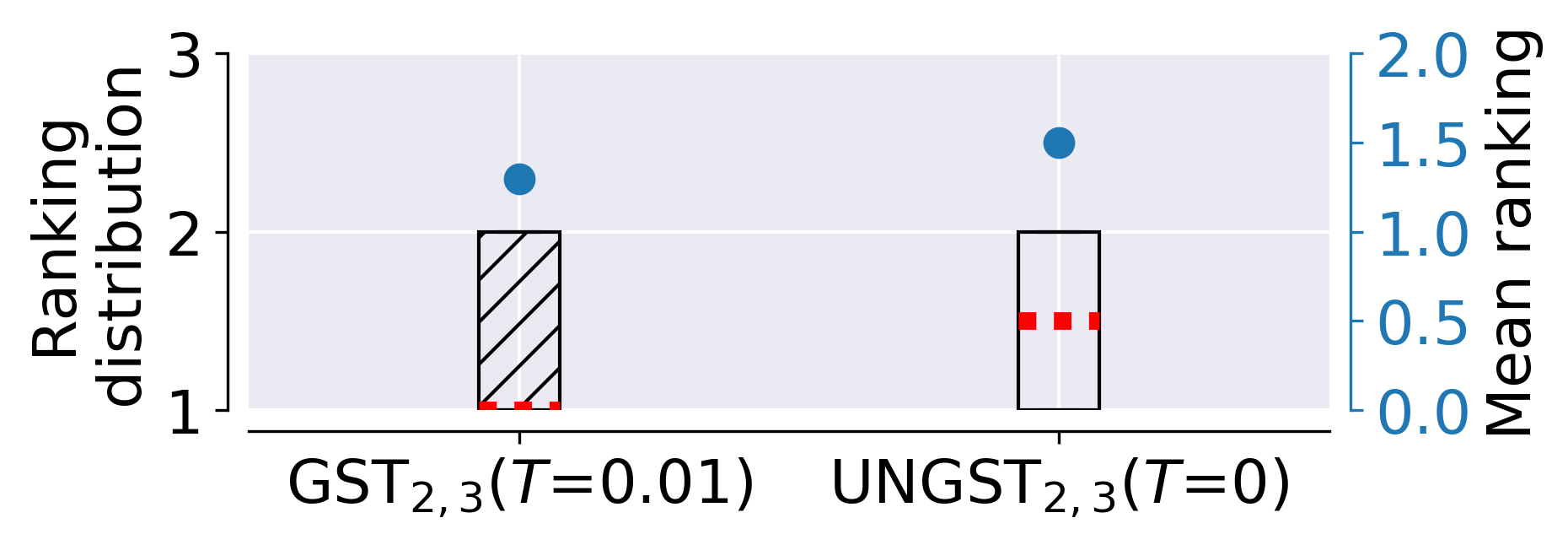}
  \caption{The ranking distribution and mean ranking of GST$_{2,3}$($T=0.01$) and UNGST$_{2,3}$($T=0$) (the unnormalized version by removing $\frac{1}{|L_{l}(u, \mathcal{G})|}$ from Eq.~( \ref{Eq_Distance}), summarized over six structural queries for Glo\_ERA5SP.
  GST is highlighted with hatches in boxplots and the median of the ranking distribution is shown with red dash lines.
  This figure indicates the necessity of including a normalization factor for better performance.}
  \label{Fig:Ranking_With_UN}
\end{figure}
\begin{figure}[!htbp]
  \centering
  \includegraphics[width=0.5\textwidth]{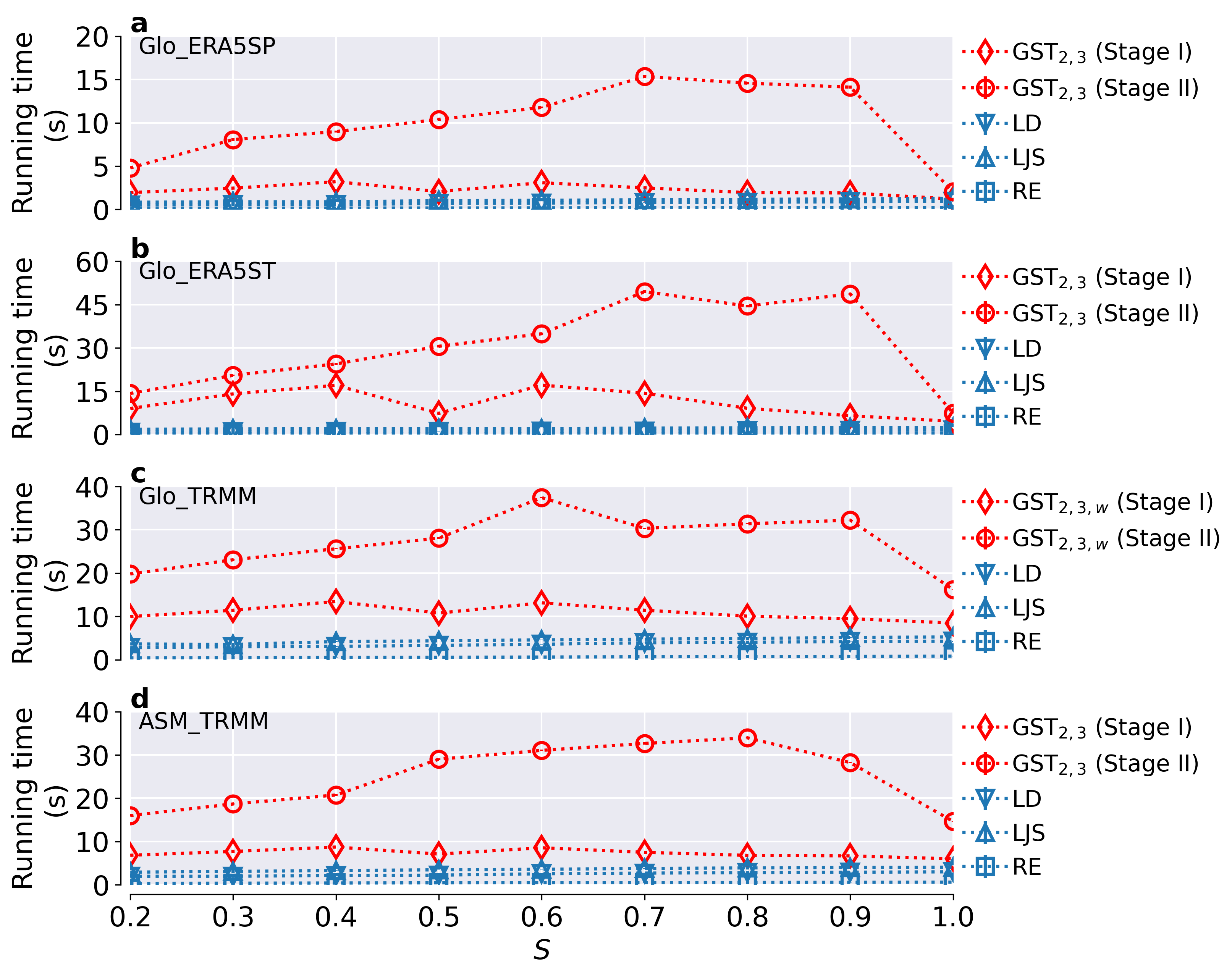}
  \caption{The running times of GST, LD, LJS and RE.
  For each network, GST chooses the best preservation scenario based on Figure~\ref{Fig:Ranking_all}.
  (a) Glo\_ERA5SP with GST$_{2,3}$($T=0.01$). (b) Glo\_ERA5ST with GST$_{2,3}$($T=0.01$). (c) Glo\_TRMM with GST$_{2,3,w}$($T=0.01$). (d) ASM\_TRMM with GST$_{2,3}$($T=0.01$).
  Clearly, Stage II dominates the running time of GST.
  This figure indicates that in spite of a higher running time, GST can be applied to large-scale networks.}
  \label{Fig:Running_Time}
\end{figure}

As for the convergence and cumulative time of GST, due to limited space, we show the result for Glo\_ERA5SP as an example in Figure~\ref{Fig:ERA5_SP_23_23w_and_T}; results for the other three networks are similar to this one.
When the tolerance $T=0.01$ is empirically given, the convergence of $\overline{\Delta}_{2,3,w}(G')$ strictly follows the convergence trajectories of $T=0$, as it should be.
The final running times of GST$_{2,3}$($T=0.01$) and GST$_{2,3,w}$($T=0.01$) (blue circles) are at least 4 times faster than that of GST$_{2,3}$($T=0$) and GST$_{2,3,w}$($T=0$) (blue triangles).
More importantly, the qualities of the final sparse structure by $T=0.01$ (the last black circles) are still quite close to those of $T=0$ (the last black triangles).

\subsection{Complex property preservation}\label{Sec:Complex property preservation}

As for property queries,
how to compare the similarity estimates obtained for different queries is not obvious, as mentioned in Section~\ref{Sec:Experimental settings}.
We therefore give such a similarity result only for Glo\_ERA5SP as an example in Figure~\ref{Fig:ERA5_SP_23}, and focus on overall rankings in Figure~\ref{Fig:Ranking_all}.
In Figure~\ref{Fig:ERA5_SP_23}, GST$_{2,3}$($T=0.01$) and GST$_{2,3,w}$($T=0.01$) are quite good at preserving degrees (see Figure~\ref{Fig:ERA5_SP_23}c), which can be expected due to the explicit preservation of scaled degrees.
Although~\citet{hamann2016structurepreserving} concluded that LD is best for preserving the overall connectivity of a network, we here see from Figures~\ref{Fig:ERA5_SP_23}a and~\ref{Fig:ERA5_SP_23}b that LJS is even better.
This should be due to different network structures in different domains.
They use mostly social networks, while we focus on functional climate networks.
Another noteworthy point is that for a given scaling factor $S$, only similarity estimates of community structure show a slightly larger variance.
This suggests the stability of all these sampling methods applicable for practical scenarios.
Still, comparing different queries in Figure~\ref{Fig:ERA5_SP_23} is not conclusive due to
the diverse performance of the different methods.
We thus summarize Figure~\ref{Fig:ERA5_SP_23} in Figure~\ref{Fig:Ranking_all}a by using their rankings.

In Figure~\ref{Fig:Ranking_all}a, GST$_{2,3}$($T=0.01$) ranks first in the comparisons of both median (red dash lines) and mean (blue dots) rankings.
From Figure~\ref{Fig:Ranking_all}, one can conclude that, for a given $\mathcal{G}$, a good performance of GST$_{2,3}$($T=0.01$) does not guarantee that GST$_{2,3,w}$($T=0.01$) also has a similarly good performance.
For example, GST$_{2,3}$($T=0.01$) works better for Glo\_ERA5SP, Glo\_ERA5ST, and ASM\_TRMM, while GST$_{2,3,w}$($T=0.01$) is better for Glo\_TRMM and ASM\_TRMM.
We conjecture that this is due to the diversity of different network structures, as we expected in Secs.~\ref{Sec:Preliminaries} and~\ref{Sec:GST}.
Nonetheless, either one of our two methods is always the best.
Thus, preserving both scaled degrees and 3-node subgraphs yields a sparser graph that better preserves complex properties overall. This answers \textbf{Q2}.

We also compare GST$_{2,3}$($T=0.01$) with UNGST$_{2,3}$($T=0$) to verify the necessity of including a normalization factor $\frac{1}{|L_{l}(u, \mathcal{G})|}$ in Eq.~(\ref{Eq_Distance}), where
 UNGST$_{2,3}$($T=0$) is an unnormalized version of GST by removing from Eqs.~(\ref{Eq_Distance}) and~(\ref{Eq_Gain_Individual}) this normalization factor.
The same estimation process based on the above six property queries is adopted and the summarized rankings are shown in Figure~\ref{Fig:Ranking_With_UN}.
UNGST$_{2,3}$($T=0$) proceeds until the final convergence instead of early convergence, since $T=0$.
GST$_{2,3}$($T=0.01$) still generates sparse networks with a higher similarity to the original Glo\_ERA5SP properties.

\subsection{Running times}\label{Sec:Running time consumption}
To answer \textbf{Q3}, we compare the running times of GST, LD, LJS and RE in Figure~\ref{Fig:Running_Time}.
Taking $\mathcal{G}=$ Glo\_ERA5SP as an example, GST$_{2,3}$($T=0.01$) is chosen as the sampling method based on Figure~\ref{Fig:Ranking_all}a.
For each given scaling factor $S$, GST$_{2,3}$($T=0.01$) generates 100 sparse subgraphs $G^*$.
We calculate the average and deviation of running time for both stages of Algorithm~\ref{Alg:GST}.
The edge ratio between $G^*$ and $\mathcal{G}$ is used for the initialization of LD, LJS, and RE, further to obtain the corresponding running time.

According to~\cite{hamann2016structurepreserving}, the running times of LD and LJS are slightly slower than RE, which only takes linear time in the number of edges.
The running time of GST mainly depends on the number of iterations $r$ in Stage II, even with a tolerance factor $T$ included for early termination.
In Figure~\ref{Fig:Running_Time}, GST is therefore roughly 19, 12, and 90 times slower than LD, LJS, and RE, respectively.
Nonetheless, GST is still applicable to large-scale networks.

\section{Conclusion}\label{Sec:Conclusion}
In summary, we proposed a different perspective (by preserving scaled local node characteristics) from the general filtering-based sampling methods for network sparsification.
Our empirical studies {on functional climate networks} verify that the proposed method generates sparse subgraphs
that preserve the overall similarity to the original network in a considerably better way.

{As future work, we will further study the robustness of this method in other network application scenarios as well as for synthetic data.}
Which preservation of $3$-node subgraphs ($l=2,3$ or $l=2,3,w$) one should choose for
best results on {a wide range of} unknown data sets remains
{as another issue not fully settled yet}.

\section*{Acknowledgements}
We would like to thank Panos Parchas for data sharing.
Z.S. was funded by the China Scholarship Council (CSC) scholarship.
J.K. was supported by the Federal Ministry of Education and Research (BMBF) grant No. 01LP1902J (climXtreme).
H.M. was partially supported by German Research Foundation (DFG) grant ME-3619/4-1 (ALMACOM).

\bibliographystyle{IEEEtranN}
\footnotesize
\bibliography{ASONAM_2022}

\begin{thebibliography}{24}
\providecommand{\natexlab}[1]{#1}
\providecommand{\url}[1]{#1}
\csname url@samestyle\endcsname
\providecommand{\newblock}{\relax}
\providecommand{\bibinfo}[2]{#2}
\providecommand{\BIBentrySTDinterwordspacing}{\spaceskip=0pt\relax}
\providecommand{\BIBentryALTinterwordstretchfactor}{4}
\providecommand{\BIBentryALTinterwordspacing}{\spaceskip=\fontdimen2\font plus
\BIBentryALTinterwordstretchfactor\fontdimen3\font minus
  \fontdimen4\font\relax}
\providecommand{\BIBforeignlanguage}[2]{{%
\expandafter\ifx\csname l@#1\endcsname\relax
\typeout{** WARNING: IEEEtranN.bst: No hyphenation pattern has been}%
\typeout{** loaded for the language `#1'. Using the pattern for}%
\typeout{** the default language instead.}%
\else
\language=\csname l@#1\endcsname
\fi
#2}}
\providecommand{\BIBdecl}{\relax}
\BIBdecl

\bibitem[Su et~al.(2022)Su, Meyerhenke, and Kurths]{su2022climatic}
Z.~Su, H.~Meyerhenke, and J.~Kurths, ``The climatic interdependence of
  extreme-rainfall events around the globe,'' \emph{Chaos: An Interdisciplinary
  Journal of Nonlinear Science}, vol.~32, no.~4, p. 043126, 2022.

\bibitem[Hamann et~al.(2016)Hamann, Lindner, Meyerhenke, Staudt, and
  Wagner]{hamann2016structurepreserving}
M.~Hamann, G.~Lindner, H.~Meyerhenke, C.~L. Staudt, and D.~Wagner,
  ``Structure-preserving sparsification methods for social networks,''
  \emph{Social Network Analysis and Mining}, vol.~6, no.~1, p.~22, 2016.

\bibitem[Batson et~al.(2013)Batson, Spielman, Srivastava, and
  Teng]{batson2013spectral}
J.~Batson, D.~A. Spielman, N.~Srivastava, and S.-H. Teng, ``Spectral
  sparsification of graphs: Theory and algorithms,'' \emph{Communications of
  the ACM}, vol.~56, no.~8, pp. 87--94, 2013.

\bibitem[Sadhanala et~al.(2016)Sadhanala, Wang, and
  Tibshirani]{sadhanala2016graph}
V.~Sadhanala, Y.-X. Wang, and R.~Tibshirani, ``Graph {{Sparsification
  Approaches}} for {{Laplacian Smoothing}},'' in \emph{Proceedings of the 19th
  {{International Conference}} on {{Artificial Intelligence}} and
  {{Statistics}}}.\hskip 1em plus 0.5em minus 0.4em\relax {PMLR}, 2016, pp.
  1250--1259.

\bibitem[Mahadevan et~al.(2006)Mahadevan, Krioukov, Fall, and
  Vahdat]{mahadevan2006systematica}
P.~Mahadevan, D.~Krioukov, K.~Fall, and A.~Vahdat, ``Systematic topology
  analysis and generation using degree correlations,'' \emph{ACM SIGCOMM
  Computer Communication Review}, vol.~36, no.~4, pp. 135--146, 2006.

\bibitem[Orsini et~al.(2015)Orsini, Dankulov, {Colomer-de-Sim{\'o}n},
  Jamakovic, Mahadevan, Vahdat, Bassler, Toroczkai, Bogu{\~n}{\'a}, Caldarelli,
  Fortunato, and Krioukov]{orsini2015quantifying}
C.~Orsini, M.~M. Dankulov, P.~{Colomer-de-Sim{\'o}n}, A.~Jamakovic,
  P.~Mahadevan, A.~Vahdat, K.~E. Bassler, Z.~Toroczkai, M.~Bogu{\~n}{\'a},
  G.~Caldarelli, S.~Fortunato, and D.~Krioukov, ``Quantifying randomness in
  real networks,'' \emph{Nature Communications}, vol.~6, no.~1, p. 8627, 2015.

\bibitem[Benzi and Klymko(2015)]{benzi2015limiting}
M.~Benzi and C.~Klymko, ``On the {{Limiting Behavior}} of {{Parameter-Dependent
  Network Centrality Measures}},'' \emph{SIAM Journal on Matrix Analysis and
  Applications}, vol.~36, no.~2, pp. 686--706, 2015.

\bibitem[Zeng et~al.(2021)Zeng, Song, and Ge]{zeng2021selective}
Y.~Zeng, C.~Song, and T.~Ge, ``Selective {{Edge Shedding}} in {{Large Graphs
  Under Resource Constraints}},'' in \emph{2021 {{IEEE}} 37th {{International
  Conference}} on {{Data Engineering}} ({{ICDE}})}, 2021, pp. 2057--2062.

\bibitem[Zeng et~al.(2022)Zeng, Song, Ge, and Zhang]{zeng2022reductiona}
Y.~Zeng, C.~Song, T.~Ge, and Y.~Zhang, ``Reduction of large-scale graphs:
  {{Effective}} edge shedding at a controllable ratio under resource
  constraints,'' \emph{Knowledge-Based Systems}, vol. 240, p. 108126, 2022.

\bibitem[Parchas et~al.(2014)Parchas, Gullo, Papadias, and
  Bonchi]{parchas2014pursuit}
P.~Parchas, F.~Gullo, D.~Papadias, and F.~Bonchi, ``The pursuit of a good
  possible world: Extracting representative instances of uncertain graphs,'' in
  \emph{Proceedings of the 2014 {{ACM SIGMOD International Conference}} on
  {{Management}} of {{Data}}}, ser. {{SIGMOD}} '14.\hskip 1em plus 0.5em minus
  0.4em\relax {New York, NY, USA}: {Association for Computing Machinery}, 2014,
  pp. 967--978.

\bibitem[Parchas et~al.(2015)Parchas, Gullo, Papadias, and
  Bonchi]{parchas2015uncertain}
------, ``Uncertain {{Graph Processing}} through {{Representative
  Instances}},'' \emph{ACM Transactions on Database Systems}, vol.~40, no.~3,
  pp. 20:1--20:39, 2015.

\bibitem[Song et~al.(2016)Song, Zou, and Liu]{song2016trianglebased}
S.~Song, Z.~Zou, and K.~Liu, ``Triangle-{{Based Representative Possible
  Worlds}} of {{Uncertain Graphs}},'' in \emph{Database {{Systems}} for
  {{Advanced Applications}}}, ser. Lecture {{Notes}} in {{Computer Science}},
  S.~B. Navathe, W.~Wu, S.~Shekhar, X.~Du, S.~X. Wang, and H.~Xiong, Eds.\hskip
  1em plus 0.5em minus 0.4em\relax {Cham}: {Springer International Publishing},
  2016, pp. 283--298.

\bibitem[Bonchi et~al.(2014)Bonchi, Gullo, Kaltenbrunner, and
  Volkovich]{bonchi2014core}
F.~Bonchi, F.~Gullo, A.~Kaltenbrunner, and Y.~Volkovich, ``Core decomposition
  of uncertain graphs,'' in \emph{Proceedings of the 20th {{ACM SIGKDD}}
  International Conference on {{Knowledge}} Discovery and Data Mining}, ser.
  {{KDD}} '14.\hskip 1em plus 0.5em minus 0.4em\relax {New York, NY, USA}:
  {Association for Computing Machinery}, 2014, pp. 1316--1325.

\bibitem[Micciancio(2001)]{micciancio2001hardness}
D.~Micciancio, ``The hardness of the closest vector problem with
  preprocessing,'' \emph{IEEE Transactions on Information Theory}, vol.~47,
  no.~3, pp. 1212--1215, 2001.

\bibitem[Monderer and Shapley(1996)]{monderer1996potential}
D.~Monderer and L.~S. Shapley, ``Potential {{Games}},'' \emph{Games and
  Economic Behavior}, vol.~14, no.~1, pp. 124--143, 1996.

\bibitem[Ortmann and Brandes(2013)]{ortmann2013triangle}
M.~Ortmann and U.~Brandes, ``Triangle {{Listing Algorithms}}: {{Back}} from the
  {{Diversion}},'' in \emph{2014 {{Proceedings}} of the {{Meeting}} on
  {{Algorithm Engineering}} and {{Experiments}} ({{ALENEX}})}, ser.
  Proceedings.\hskip 1em plus 0.5em minus 0.4em\relax {Society for Industrial
  and Applied Mathematics}, 2013, pp. 1--8.

\bibitem[Hersbach et~al.(2020)Hersbach, Bell, Berrisford, Hirahara,
  Hor{\'a}nyi, {Mu{\~n}oz-Sabater}, Nicolas, Peubey, Radu, Schepers, Simmons,
  Soci, Abdalla, Abellan, Balsamo, Bechtold, Biavati, Bidlot, Bonavita,
  De~Chiara, Dahlgren, Dee, Diamantakis, Dragani, Flemming, Forbes, Fuentes,
  Geer, Haimberger, Healy, Hogan, H{\'o}lm, Janiskov{\'a}, Keeley, Laloyaux,
  Lopez, Lupu, Radnoti, {de Rosnay}, Rozum, Vamborg, Villaume, and
  Th{\'e}paut]{hersbach2020era5b}
H.~Hersbach, B.~Bell, P.~Berrisford, S.~Hirahara, A.~Hor{\'a}nyi,
  J.~{Mu{\~n}oz-Sabater}, J.~Nicolas, C.~Peubey, R.~Radu, D.~Schepers,
  A.~Simmons, C.~Soci, S.~Abdalla, X.~Abellan, G.~Balsamo, P.~Bechtold,
  G.~Biavati, J.~Bidlot, M.~Bonavita, G.~De~Chiara, P.~Dahlgren, D.~Dee,
  M.~Diamantakis, R.~Dragani, J.~Flemming, R.~Forbes, M.~Fuentes, A.~Geer,
  L.~Haimberger, S.~Healy, R.~J. Hogan, E.~H{\'o}lm, M.~Janiskov{\'a},
  S.~Keeley, P.~Laloyaux, P.~Lopez, C.~Lupu, G.~Radnoti, P.~{de Rosnay},
  I.~Rozum, F.~Vamborg, S.~Villaume, and J.-N. Th{\'e}paut, ``The {{ERA5}}
  global reanalysis,'' \emph{Quarterly Journal of the Royal Meteorological
  Society}, vol. 146, no. 730, pp. 1999--2049, 2020.

\bibitem[Gupta et~al.(2021)Gupta, Boers, Pappenberger, and
  Kurths]{gupta2021complex}
S.~Gupta, N.~Boers, F.~Pappenberger, and J.~Kurths, ``Complex network approach
  for detecting tropical cyclones,'' \emph{Climate Dynamics}, vol.~57, no.~11,
  pp. 3355--3364, 2021.

\bibitem[Huffman et~al.(2007)Huffman, Bolvin, Nelkin, Wolff, Adler, Gu, Hong,
  Bowman, and Stocker]{huffman2007trmmb}
G.~J. Huffman, D.~T. Bolvin, E.~J. Nelkin, D.~B. Wolff, R.~F. Adler, G.~Gu,
  Y.~Hong, K.~P. Bowman, and E.~F. Stocker, ``The {{TRMM Multisatellite
  Precipitation Analysis}} ({{TMPA}}): {{Quasi-Global}}, {{Multiyear}},
  {{Combined-Sensor Precipitation Estimates}} at {{Fine Scales}},''
  \emph{Journal of Hydrometeorology}, vol.~8, no.~1, pp. 38--55, 2007.

\bibitem[Quian~Quiroga et~al.(2002)Quian~Quiroga, Kreuz, and
  Grassberger]{quianquiroga2002eventb}
R.~Quian~Quiroga, T.~Kreuz, and P.~Grassberger, ``Event synchronization: {{A}}
  simple and fast method to measure synchronicity and time delay patterns,''
  \emph{Physical Review E}, vol.~66, no.~4, p. 041904, 2002.

\bibitem[Satuluri et~al.(2011)Satuluri, Parthasarathy, and
  Ruan]{satuluri2011local}
V.~Satuluri, S.~Parthasarathy, and Y.~Ruan, ``Local graph sparsification for
  scalable clustering,'' in \emph{Proceedings of the 2011 {{ACM SIGMOD
  International Conference}} on {{Management}} of Data}, ser. {{SIGMOD}}
  '11.\hskip 1em plus 0.5em minus 0.4em\relax {New York, NY, USA}: {Association
  for Computing Machinery}, 2011, pp. 721--732.

\bibitem[Staudt et~al.(2016)Staudt, Sazonovs, and
  Meyerhenke]{staudt2016networkita}
C.~L. Staudt, A.~Sazonovs, and H.~Meyerhenke, ``{{NetworKit}}: {{A}} tool suite
  for large-scale complex network analysis,'' \emph{Network Science}, vol.~4,
  no.~4, pp. 508--530, 2016.

\bibitem[Vinh et~al.(2010)Vinh, Epps, and Bailey]{vinh2010information}
N.~X. Vinh, J.~Epps, and J.~Bailey, ``Information {{Theoretic Measures}} for
  {{Clusterings Comparison}}: {{Variants}}, {{Properties}}, {{Normalization}}
  and {{Correction}} for {{Chance}},'' \emph{Journal of Machine Learning
  Research}, vol.~11, no.~95, pp. 2837--2854, 2010.

\bibitem[Staudt and Meyerhenke(2016)]{staudt2016engineeringa}
C.~L. Staudt and H.~Meyerhenke, ``Engineering {{Parallel Algorithms}} for
  {{Community Detection}} in {{Massive Networks}},'' \emph{IEEE Transactions on
  Parallel and Distributed Systems}, vol.~27, no.~1, pp. 171--184, 2016.

\end{thebibliography}

\end{document}